# Evolving Testing Scenario Generation Method and Intelligence Evaluation Framework for Automated Vehicles

Yining Ma[1], Wei Jiang[1], Lingtong Zhang[1], Junyi Chen[1,*], Hong Wang[2], Chen Lv[3], Xuesong Wang[4, 5], Lu Xiong[1]

[1] School of Automotive Studies, Tongji University, Shanghai 201804, China
[2] School of Vehicle and Mobility, Tsinghua University, Beijing 100084, China.
[3] School of Mechanical and Aerospace Engineering, Nanyang Technological University, Singapore 639798, Singapore
[4] The Key Laboratory of Road and Traffic Engineering, Ministry of Education, China
[5] College of Transportation Engineering, Tongji University, Shanghai 201804, China

*Abstract* — Interaction between the background vehicles (BVs) and automated vehicles (AVs) in scenario-based testing plays a critical role in evaluating the intelligence of the AVs. Current testing scenarios typically employ predefined or scripted BVs, which inadequately reflect the complexity of human-like social behaviors in real-world driving scenarios, and also lack a systematic metric for evaluating the comprehensive intelligence of AVs. Therefore, this paper proposes an evolving scenario generation method that utilizes deep reinforcement learning (DRL) to create human-like BVs for testing and intelligence evaluation of AVs. Firstly, a class of driver models with human-like competitive, cooperative, and mutual driving motivations is designed. Then, utilizing an improved "level-k" training procedure, the three distinct driver models acquire game-based interactive driving policies. And these models are assigned to BVs for generating evolving scenarios in which all BVs can interact continuously and evolve diverse contents. Next, a framework including safety, driving efficiency, and interaction utility are presented to evaluate and quantify the intelligence performance of 3 systems under test (SUTs), indicating the effectiveness of the evolving scenario for intelligence testing. Finally, the complexity and fidelity of the proposed evolving testing scenario are validated. The results demonstrate that the proposed evolving scenario exhibits the highest level of complexity compared to other baseline scenarios and has more than 85% similarity to naturalistic driving data. This highlights the potential of the proposed method to facilitate the development and evaluation of high-level AVs in a realistic and challenging environment.

*Index Terms* — Automated vehicle, scenario-based testing, human-like social driver model, intelligence evaluation, deep reinforcement learning.

## I. INTRODUCTION

Autonomous driving technology has entered into a rapid development phase. High-level automated vehicles (AVs) should ensure driving safety, high efficiency, comfort, and reasonable interactivity[1]. Road testing is one available way to achieve thorough reliability validation, but this approach is time-consuming and expensive[2][3]. An alternative approach is scenario-based testing, which effectively verifies and validates autonomous driving systems[4][5]. The main approaches for generating scenarios in the current research field can be broadly classified into two categories:

- The analysis-based predefined scenarios, where the traffic participants don't have interaction-aware and only the system under test (SUT) can react to other traffic participants, it can be called one-way interaction scenarios[6].

*Corresponding author. Email: chenjunyi@tongji.edu.cn

- The evolving scenarios are trained based on machine learning, where the traffic participants are implemented with intelligent driver models or controlled by other humans in a multi-player game fashion, it can be called a bidirectional or multiple interaction scenario[7]. The background vehicles (BVs) evolve different interactions based on each behavior of the SUT, so each future time step is difficult to specify and predict.

Compared with evolving scenarios, there remain several limitations in the predefined scenario: 1) Traffic environment is mainly based on scripted and episodic driving scenarios, in which the actions of traffic participants are predefined in response to the SUT. Thus, the hard coding driving behaviors would generate some single interaction category simulation environments, which might cause low effective testing. 2) Human-like social driving policies of BVs in the predefined scenario are lacking, which may lead to poor gaming and unrealistic interaction between BVs and SUT. 3) Predefined scenarios usually have a targeted goal in testing, such as for a certain metric of the SUT, and are less often used for multi-objective testing. The intelligence of the SUT is a combination of several metrics, so it isn't easy to satisfy the intelligence-oriented testing of the SUT based on predefined testing scenarios.

Therefore, how to design and generate an evolving scenario with high testing efficiency, complexity, and fidelity; and which metrics can accurately evaluate the intelligence of the SUT within this testing scenario become two key issues. In light of this, we build a set of human-like social driver models to generate evolving testing scenarios and propose a multi-dimensional metric framework geared towards the intelligence evaluation of the decision-making system (SUT). We first designed a novel class of human-like social driver models, which be trained game-like driving policies through "level-k" theory and TD3 algorithm, and these models are then adopted as BVs in generating evolving scenarios. Especially by shaping the reward functions, the driving models are endowed with competitive, mutual, and cooperative game-like driving policies. Then, a framework is proposed for the intelligence evaluation of the SUT, and a few innovative metrics are designed. We validated the effectiveness of the generated testing scenarios based on the intelligence evaluation framework. Finally, we validated the complexity and fidelity of the evolving scenarios. The whole framework diagram of this paper is shown in Fig. 1, and the specific contributions are listed as follows:

1) A type of human-like social driver model is proposed that can be used to generate evolving testing scenarios with extensive spatial-temporal scale interactions and high degrees of uncertainty. These driving models are trained to acquire intelligence and develop three distinct game-like interaction policies: cooperative, mutual, and competitive. 2) We present a quantitative evaluation framework that includes multiple metrics: safety, driving efficiency, and interaction utility, which can differentiate the disparities in the intelligence of diverse SUTs in evolving testing scenarios. 3) The driver models employ a hierarchical approach, whereby higher-level decision-making is assigned human-like game policies, and lower-level control is entrusted to a high-precision dynamics model. 4) A comprehensive validation and comparison of the generated evolving testing scenarios were realized.

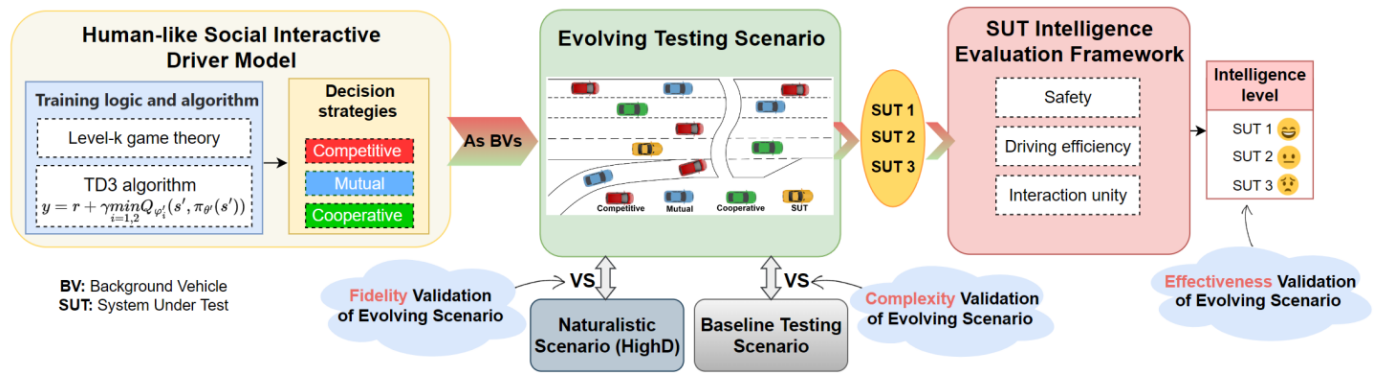

Fig. 1. The framework diagram of this paper

The paper is structured as follows: In Sec.II, the related works are introduced and analyzed. In Sec.III, the methodology, including the scheme of the driver models, the training algorithm and the environment, are presented. Sec.IV demonstrates the training procedure, including the simulation platform, the process of generating human-like gaming driving policies of driver models, and the training results. In Sec.V, the intelligence of different SUTs is evaluated, and the evolving testing scenarios are validated. In Sec.VI, conclusions are summarized, and future research priorities are derived.

## II. RELATED WORK

### A. Generation of Evolving Testing Scenario

Evolving testing scenario allows the BVs autonomously take driving actions in response to the SUT. To increase the effectiveness and efficiency of simulation testing, three primary methods are commonly employed to generate evolving driving scenarios.

The first type is heuristic methods, including Bayesian optimization and hyperparameter search[8][9]. Bayesian optimization is introduced to generate possible driving policy parameters, which could cause an adversarial driving scenario in a simulator[10][11]. Hyperparameter search methods are proposed to find dangerous driving scenarios against SUT weaknesses[12]. The testing scenarios generated by this type of method have better testing efficiency but require a given parameter range[13]. Moreover, it mainly generates short scenario segments with limited traffic participants, so it cannot evolve diverse content.

The second type is based on imitation learning (IL) and deep learning (DL)[14][15]. Inverse reinforcement learning, as a representative method of IL, trains human-like driving models by extracting typical features from naturalistic driving data (NDD). For example, in [16][17][18], a detailed trajectory highly similar to ground-truth human driving was generated through personification modeling. Some studies have utilized DL algorithms such as GAN and GRU to decode and encode generalized trajectories with different interaction abilities. These trajectories are then clustered and used to generate testing scenarios[19][20]. Since the data used to generate these scenarios through IL and DL comes from the real world, the scenarios possess high fidelity. However, these scenarios lack challenge, as safety-critical data is scarce in NDD[21]. Furthermore, the trajectories in the scenarios are generated through end-to-end methods, which also lack interpretability [22].

The third class of methods for generating evolving scenarios is based on adversarial reinforcement learning (RL). A representative study from Feng's team[23][24]. They trained the BVs to execute adversarial maneuvers appropriately to increase the scenario's challenge and proved the high unbiasedness between the testing scenario and the naturalistic traffic environment[25]. In the latest study, they proposed a dense deep-reinforcement-learning (D2RL) approach, which can observably reduce the variance of the policy gradient estimation without loss of unbiasedness[26]. The evolving scenarios generated through deep reinforcement learning (DRL) (or RL) show great potential for their ability to balance fidelity and testing efficiency.

The methods above share a common feature, which is to learn adversarial maneuvers or behaviors directly from the statistical results of NDD when building driver models. The driver model learns the underlying driving behaviors of human drivers, including lane changing, following, cutting, and other behaviors, which are then used to generate testing scenarios. However, these methods lack the higher-level motivational design and causal analysis in the driver model.

In contrast to the aforementioned studies, we began our modeling process at the higher-level driver motivation. We designed three motivations to build the driver model, which are competitive, cooperative, and mutual. The driver models learn these interactive motivations through DRL training and derive realistic driving behaviors. The idea of beginning the modeling process with motivation originates from The Motivation- Ability-Opportunity (MAO) Model, a theoretical framework used in psychology to explain human behavior[27]. The MAO model suggests that behavior results from the interaction between an individual's motivation, ability, and opportunity to engage in the behavior. First, at the motivation level, we designed three driving policies (competitive, mutual, and cooperative). Then, at the ability level, we trained the three models using game theory and the TD3 algorithm to learn the diverse behaviors. Finally, at the opportunity level, we combined the three types of driver models and allow them to interact to generate the evolving scenarios. In these scenarios, different interactions ultimately result in different driving behaviors.

## B. Characteristics of Social Driving Behaviors

The characteristics of social driving behavior are diverse and requirements definition and classification. Some studies define social driving behavior through theoretical analysis or a knowledge-driven approach[28]. For example, Wang et al. provided a quantifiable definition of social interaction behavior by combining social psychology and traffic psychology[6]. A study[29] combined social psychology and game theory to define and model socially compliant driving behavior. For the classification of social driving behavior, most studies employ a parameter clustering approach or unsupervised/semi-supervised learning to classify and analyze driving behavior characteristics[30]. For instance, driving behaviors are commonly categorized into aggressive, moderate, and conservative types based on statistical clustering of parameters such as Time-To-Collision (TTC), lane change frequency, lateral and longitudinal acceleration, etc.[31]. These categories of driving behavior emphasize the driver's individual attributes and ignore their interaction with other vehicles, which does not fully reflect their social nature.

In this study, we emphasize the interaction between driving behaviors and model them through different interaction motivations to make driving behavior more social. Moreover, we have, for the first time, applied these socially-oriented driver models (competitive, neutral, and cooperative) to the design of BVs in testing scenarios. These diverse driving behaviors make testing scenarios unpredictable and challenging.

## C. Intelligence Evaluation

Most previous studies on AV evaluation tend to focus on validating that basic functionality or main performance meets the requirements, such as safety, efficiency, comfort, etc. while lacking validation of AVs' overall and comprehensive intelligence[32]. For example, the various competitions, such as DARPA[34] and the European land robot test[35], are most task-driven. The evaluation metrics of the competitions are relatively individual, resulting in the tested vehicles being unable to obtain the intelligence level grading or intelligence degree quantification results corresponding to the testing results.

In recent years, some comprehensive evaluation frameworks based on multiple metrics are gradually becoming mainstream due to the rapid development of AVs[36][37]. Such as Zhao et al. proposed a comprehensive evaluation approach with crash probability, collision probability, and serious injury probability as the main indicators in the acceleration test[38]. In addition, a new concept of the "driving risk field" is proposed by [39], which includes the kinetic field, potential field, and behavioral field.

In conclusion, certain evaluation metrics prioritize specific aspects while disregarding integrated intelligence. Other evaluations consider the role of the driving task and the environment, but do not consider the effect of the interactive behavior of the SUT on the BV. In other words, the existing evaluation methods do not explicitly quantify the ability of the SUT to make adaptive decisions and actions in an interactive environment[40]. However, the ability of the SUT to interact with the BV is an important manifestation of its intelligence. Therefore, current evaluation studies lack a comprehensive framework for evaluating the intelligence of SUTs.

## III. METHODOLOGY

### A. Background

An RL problem can be interpreted as an agent learning via interaction with its environment driven by a feedback signal (reward)[41], and the environment reinforces the agent to determine better actions to promote the learning process[42][43]. Usually, such interaction can be modeled by a Markov Decision Process (MDP). A fully observed MDP question is generally defined by 5-tuple $(\mathcal{S}, \mathcal{A}, P_{ss'}, \mathcal{R}, \gamma)$, in which $\mathcal{S}$ indicates the state space, $\mathcal{A}$ is the action space, $P_{ss'}$ is the transition possibility of the environment from state $s$ to state $s'$. $\mathcal{R}$ is the delayed reward function, $\gamma$ is the discount factor. Policy $\pi$ is the reflection from state $s$ to action $a$. Assume the delayed reward received by the agent at time $t+1$ is $r_{t+1}$, the long-term cumulative reward at time $t$ can be formulated as $G_t = \sum_{i=0}^{\infty} \gamma^i r_{t+i+1}$. Therefore, the goal of the agent in RL is to maximize the expected cumulative reward $V_\pi(s) = \mathbb{E}_\pi[G_t|s_t = s]$ under policy $\pi$ at each state $s$.

For many optimal control problems, taking the proper action under different conditions is mostly a concern.

Instead of $V_\pi(s)$, state-action value function $Q^\pi(s,a)$ is defined as:

$$Q^\pi(s,a) = \mathbb{E}_\pi\left[\sum_{i=0}^{\infty} \gamma^i r_{t+i+1} \mid s_t = s, a_t = a\right] \quad (1)$$

Thus, the optimal solution of policy $\pi$ is $\pi^*$, which makes:

$$Q^{\pi^*}(s,a) = \max_\pi Q^\pi(s,a) \quad (2)$$

The method of solving this function is known as Q-learning, which uses a temporal difference error (TD-error) update algorithm at each iteration:

$$Q^\pi(s,a) \leftarrow Q^\pi(s,a) + \alpha\left[r_{t+1} + \gamma \max_{a' \in \mathcal{A}} Q^\pi(s',a') - Q^\pi(s,a)\right] \quad (3)$$

In order to execute this update, a massive table storing all state-action pairs is needed. To better match the real motion state of the vehicle and improve the accuracy of the action, we need to calculate the vehicle action space in the continuous state. Creating such a Q-learning table is infeasible for continuous state or action space. Twin Delayed Deep Deterministic Policy Gradient Algorithm (TD3)[44] is proposed to handle the problem of continuous control under high-dimension state space. TD3 is based on the Actor-Critic framework. In TD3, the Q-learning table is replaced by four critical Q-networks to form the relationship between the state-action pair and value function $Q^\pi(s,a)$, in which two of them are value networks updated by gradient decent, and others are target networks using soft update $\varphi'_{i=1,2} = \tau\varphi_{i=1,2} + (1-\tau)\varphi'_{i=1,2}$. For continuous control, the policy $\pi$ is also parameterized as a value network $\pi_\theta(a|s)$ and target network $\pi_{\theta'}(a|s)$. As a Q-learning algorithm, TD3 is still based on TD-error update only with the estimation of future state-action value function changed:

$$y = r + \gamma \min_{i=1,2} Q_{\varphi'_i}(s', \pi_{\theta'}(s')) \quad (4)$$

A typical training pseudocode of TD3 is listed as the following Table I.

Table I

TD3 Training Algorithm

| Algorithm 1 |
|---|
| **Initialization** $s, \theta, \theta', \omega_1, \omega_2, \omega_1', \omega_2'$ parameters, $\alpha, \beta_1, \beta_2$ learning rates, $\gamma, \phi$ decay factors and replay buffer $\mathcal{B}$ |
| 1: **for** $t = 1,2,\ldots T$ **do** |
| 2:     select action $a$ based on $\pi_\theta(a|s)$ added by noise $\epsilon \sim N(0,\sigma'^2)$, clip action $a$ within $[a_{min}, a_{max}]$ |
| 3:     observe $s'$, receive $r$, push trajectory $\tau = \{s,a,s'r\}$ into $\mathcal{B}$ |
| 4:     sample mini-batch $\mathcal{T}$ of trajectories $\tau = \{s,a,s'r\}$ from replay buffer $\mathcal{B}$ |
| 5:     select next action $a'$ based on $\pi_\theta(a'|s')$ added by noise $\epsilon \sim N(0,\sigma^2)$, clip action $a'$ within $[a_{min}, a_{max}]$ |
| 6:     compute TD-error: $\delta \leftarrow r + \gamma \min_{i=1,2} Q(s',a',\omega_i') - Q(s,a,\omega_i)$ |
| 7:     update: $\omega_1 \leftarrow \omega_1 + \beta_1 \delta \nabla_{\omega_1} Q(s,a,\omega_1)$ |
| 8:     update: $\omega_2 \leftarrow \omega_2 + \beta_2 \delta \nabla_{\omega_2} Q(s,a,\omega_2)$ |
| 9:     **if** $t$ mod n **do** |
| 10:       update: $\theta \leftarrow \theta + \alpha \nabla_\theta ln\pi_\theta(s',a') Q_\omega(s',a')$ |
| 11:       update: |
| 12:         $\theta' \leftarrow \theta'\phi + \theta(1-\phi)$ |
| 13:         $\omega_1' \leftarrow \omega_1'\phi + \omega_1(1-\phi)$ |
| 14:         $\omega_2' \leftarrow \omega_2'\phi + \omega_2(1-\phi)$ |
| 15:     **end if** |
| 16: **end for** |

## B. Driver Model Scheme of BV in Ramp Scenario

In highway on-ramp scenarios, there are times when the ego vehicle will inevitably interact with other drivers. An inappropriate driving decision can lead to a potential collision, which makes the on-ramp scenario a relatively challenging driving scenario. Therefore, we select the ramp to generate the testing scenario.

In this study, we proposed an intelligent driver model which is trained to generate human-like social driving policies and serves as a BV to generate the evolving testing scenario. The scheme design of the interaction between the driver model (in the training environment viewed as an ego vehicle) and the simulation environment is illustrated in Fig.2.

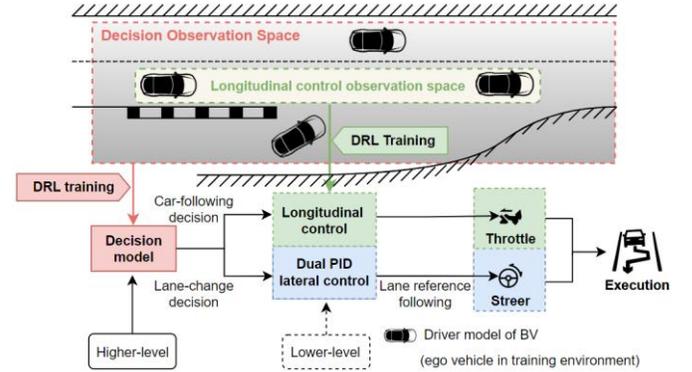

Fig.2 Scheme Design for the Driver Model

First, at the higher-level decision-making part, the decision observation space (as shown in Fig.2 grey area) of the ego vehicle is established based on the horizon field. Then, a lane-change decision can be generated by training the decision model through DRL (red area). At the lower-level of the control part, the target lane reference will change if the ego vehicle plans to execute a lane-change action. The dual PID lateral control module is utilized to realize the tracking of the target lane reference (blue area).

In the meantime, the throttle action is also trained by DRL (green area) to maintain a proper car-following gap in the established longitudinal control observation space (yellow area). Finally, the control of the ego vehicle, including throttle and steer action, will be executed to realize continuous lane-change driving behaviors based on a high-fidelity simulation platform, CARLA[45].

It should be noted that when the above designed driver models are used to generate a testing scenario in the later section, these models are referred to as the background vehicles (**BVs**). While in the training process, we refer to the designed driver model as the **ego vehicle**. The other vehicles in the training environment are called the surrounding vehicles (**SVs**).

### C. MDP Formulation

1) Observations

Based on the scheme design of the intelligent driver model, two types of observation spaces for decision-making and longitudinal control are proposed.

The first is decision observation space. A decision observation space $\Omega_d$ is a 3-dimension matrix including all the relative information of the driving state between ego vehicle and SVs in the horizon. As shown in Fig.3, the decision observation space can cover up to 5 driving lanes. The upper layer of $\Omega_d$ reveals the existence of information $P$ of SVs in the driving horizon of the ego vehicle. The rest of the decision observation space $H$ contains relative longitudinal velocity $\Delta v_s$, relative lateral velocity $\Delta v_d$ and relative yaw angle $\Delta e$. To simulate the vision in the rearview mirrors, a back view of observation is designed with a maximum distance of 20m (blue), while the maximum front view distance is 100m (green). We set the smallest grid size in the observation space is 5×3.5×1(m³), where 3.5m is the standard width of a driving lane. Therefore, the total $\Omega_d$ is divided into 500 blocks.

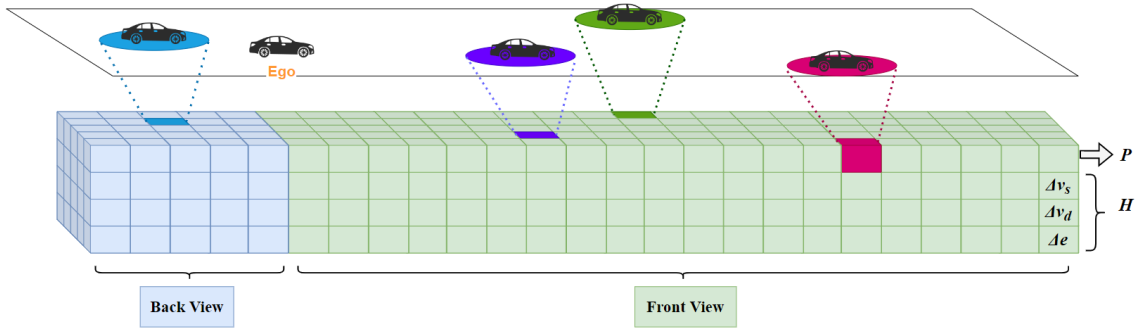

Fig.3 Decision Observation Space

The second is longitudinal car-following control observation space. In order to simulate the car-following driving behaviors of human drivers, a car-following control model is proposed using DRL. The model takes longitudinal control observation as input and generates throttle action. The control observation is represented by $s = [\Delta s, \Delta \dot{s}, \Delta \ddot{s}, \lambda]$, and the longitudinal control observation space is assumed as $o_l$, the $i_{th}$ SV is assumed as $\mathbb{v}_i$, the ego vehicle is assumed as $\mathcal{E}$, then the following functions are given:

$$\lambda = \begin{cases} 0 & if\ \mathbb{v}_i \notin o_l \\ 1 & if\ \mathbb{v}_i \in o_l \end{cases}$$
$$\Delta s = (-\mathcal{E}_s \lambda + \lambda min_{i=1,2,\dots} \mathbb{v}_{i_s})/o_{smax}$$
$$\Delta \dot{s} = (-\mathcal{E}_{\dot{s}} \lambda + \lambda min_{i=1,2,\dots} \mathbb{v}_{i_{\dot{s}}})/v_{max}$$
$$\Delta \ddot{s} = (-\mathcal{E}_{\ddot{s}} \lambda + \lambda min_{i=1,2,\dots} \mathbb{v}_{i_{\ddot{s}}})/a_{max} \quad (5)$$

where the $\Delta s$, $\Delta \dot{s}$, $\Delta \ddot{s}$ are respectively the relative distance, relative velocity, and relative acceleration between the ego vehicle and the nearest SV in the simulation. The $\lambda$ is the indicator to determine whether the $\mathbb{v}_i$ is inside the current $o_l$ or not. The $o_{smax}$ is the maximum view distance of $o_l$, the $v_{max}$ is the maximum velocity limitation, and the $a_{max}$ is the maximum acceleration limitation.

2) Actions

For lane-change action, three discrete decision actions are defined as maintain (0), left-turn (1), and right-turn (-1), where the number indicates the offset of lane reference for the lateral PID tracking control. For longitudinal car-following control, continuous throttle action is clipped to [-1, 1], where the -1 represents a full brake, and the 1 represents a full acceleration.

3) Constraints for Training

Rules are defined for more efficient training in case of undesirable driving decision actions. For example, at the beginning of decision training, the ego vehicle will make the wrong decision to drive off the road due to the trial and error mechanism in DRL. Such a decision can be manually prevented. Therefore, all lane-change actions resulting in driving off the road will be modified to a maintenance action. Moreover, a cut-in action is forbidden in case of a potential collision. It will be rectified as a maintenance action when other vehicles longitudinally exist nearby the ego vehicle within a range of 10m.

4) Rewards

Reward shaping is one of the most critical parts of RL or DRL, which should be illuminating to guide the agent to take the desired actions. A great reward function could reduce the time cost of model training and guarantee model performance.

For longitudinal car-following control training, a reward function should be formed to ensure the ego vehicle maintains the desired car-following gap with the nearest SV. Therefore, the control reward function is defined as follows:

$$r = \begin{cases} -\mathcal{C} & \text{collision happens} \\ \mu V(s') + k(V(s') - V(s)) & \text{otherwise} \end{cases} \quad (6)$$

where the $-\mathcal{C}$ is the constant punishment of collision, $V(s)$ is the current state value function and $V(s')$ is the value function of the next state. The $\mu$ and $k$ are the balance coefficients for the training agent. The state value function is defined as:

$$V(s) = -\frac{(\Delta s - g)^2}{V_{max}^2} \quad (7)$$

where the $\Delta s$ is the relative distance between the ego vehicle and the nearest observed vehicle, the $g$ is the target car-following gap, $V_{max}$ is the normalizing factor.

For the training of driving policies of the ego vehicle, we aim to generate human-like driving behaviors. Instead of individual driving behaviors, interactions happening in microscopic traffic flow are concentrated. Three game-based driving policies are proposed: competitive, mutual, and cooperative driving. The competitive driving policy is designed to maximize the individual driving utilities and diminish the driving utilities of other vehicles in the back view of the decision observation space. However, the mutual driving policy only focuses on individual driving utilities and has a sufficiently normal interaction with other vehicles. In contrast with the competitive driving policy, the cooperative driving policy is proposed to maximize the driving utilities of all vehicles in the back view of the decision observation space.

Based on the definition of different socially driving characteristics, we define decision reward function as:

$$r = r_i + r_c \quad (8)$$

where the $r_i$ indicates the individual reward, $r_c$ is the socially driving reward. More specifically, the individual reward is defined as:

$$r_i = sgn(\vartheta O_t - O_c)\frac{(\vartheta O_t - O_c)^2}{\mathcal{M}} + \mathcal{J}_{lc} + \mathcal{J}_{req} + \frac{O_c^2}{O_{max}^2} \quad (9)$$

in which the $\vartheta$ is the factor in encouraging left-hand driving behavior. The $O_t$ is the drivable area of the ego vehicle in the target driving lane, $O_c$ is the drivable area of the ego vehicle in the current driving lane. The drivable area is defined as the longitudinal distance between the ego vehicle and the nearest SV in the specified driving lane, $\mathcal{J}_{lc}$ is the punishment for lane-change actions, $\mathcal{J}_{req}$ is the punishment for the difference between the requested decision action (generated by the decision model) and the actual executed decision action (modified by rule constraints). $O_{max}$ is the maximum drivable area, which is the front view distance in $\Omega_d$, $\mathcal{M}$ is a constant for normalization.

For the mutual driving policy, the reward function $r_m$ is equal to $r_i$. For the competitive and cooperative driving policies, the reward functions are:

$$r_{com} = r_i - u\sum_{\mathbb{V}_i \in \Omega_b}\frac{(\vartheta O_c^{\mathbb{V}_i})^2}{O_{max}^2} - v\sum_{\mathbb{V}_i \in \Omega_b} \min\left(\frac{\ddot{s}_i|\ddot{s}_i|}{\ddot{s}_{max}^2}, z\right) \quad (10)$$

$$r_{coo} = r_i + m\sum_{\mathbb{V}_i \in \Omega_b}\frac{(\vartheta O_c^{\mathbb{V}_i})^2}{O_{max}^2} + n\sum_{\mathbb{V}_i \in \Omega_b} \min\left(\frac{\dot{s}_i|\dot{s}_i|}{\dot{s}_{max}^2}, y\right) \quad (11)$$

where the $O_c^{\mathbb{V}_i}$ is the drivable area of $\mathbb{V}_i$ in the current driving lane, which is defined as the longitudinal distance between $\mathbb{V}_i$ and the nearest SV in the same driving lane. $\Omega_b$ is the back view of the $\Omega_d$. The $u$, $v$, $m$ and $n$ are user-defined coefficients and the $z$ and $y$ are the reward boundaries.

## IV. TRAINING OF INTERACTIVE DRIVER MODEL

### A. Simulation Maps and Platform

This paper uses an OpenDRIVE HD map, including merging and a 4-lane dual carriageway road segment. All vehicles driving on the simulation map should obey the left-hand driving rules. According to Chinese highway traffic constraints, the maximum driving velocity is limited to 120 km/h, while the minimum driving velocity should not be less than 90 km/h. The total length of the simulation map is 2393.76m, which is long enough for simulation testing.

The high-fidelity simulator Carla based on the Unreal4 engine is introduced to precisely provide vehicle dynamics and real-time state updates to ensure accurate vehicle motions. In addition, Carla is fully capable of supporting the customized OpenDRIVE format simulation map. A snapshot of the on-ramp simulation map is shown in Fig.4.

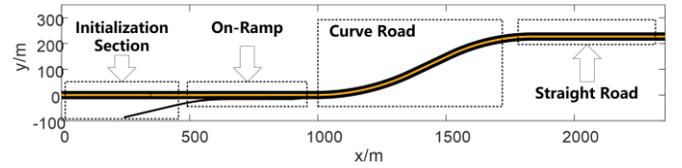

Fig.4 On-ramp Map in Carla Simulation Platform

### B. Improved Level-k Decision Training Procedure

In previous work, Nan Li[46][47] uniquely combined level-k game theory with the RL framework, where the level-k is used to construct the logic of interactions between intelligent agents, and the RL is used to evolve these interactions in time-extended scenarios. In this work, we improved their previous work by replacing the simulated interaction agents with a new intelligent driver model and using the DRL approach to evolve their interactions, as shown in Fig. 5.

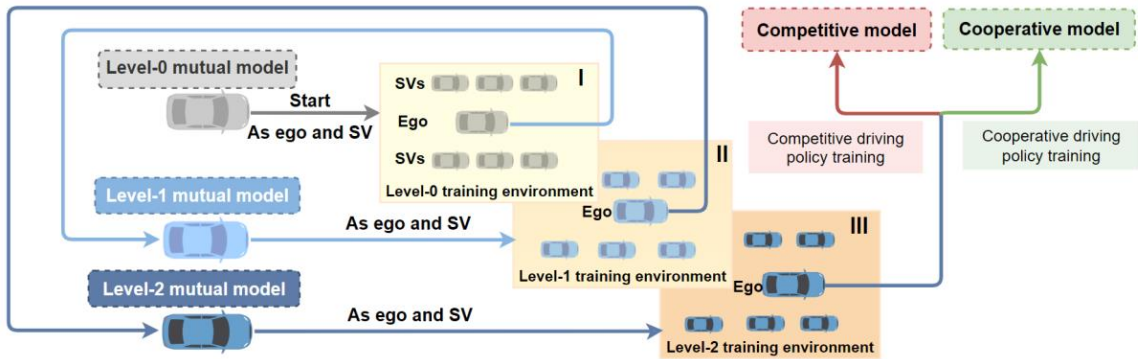

Fig.5 Level-k Decision Training Procedure

Firstly, the level-0 mutual training environment is initialized. In this environment, all the SVs except the trained ego vehicle perform the level-0 mutual driving policy, which does not actively change lanes. It means that all SVs in the training environment only perform the necessary lane-changing behavior when constrained by the road topology and pursue the lane-keeping driving policy at other times. The level-1 mutual driving decision model is obtained by training the ego vehicle in the level-0 environment. The level-1 driving model will decide on a more appropriate lane-changing or car-following behavior when interacting with the SVs. Secondly, a level-1 mutual training environment is established with the level-1 mutual driving decision model as the SV. The level-2 mutual driving decision model is obtained by training the ego vehicle in a level-1 environment. The level-2 driving model has the interactive decision capability to deal with complex lane change scenarios. Compared with the level-1 driving model, the lane-change behavior of the level-2 driving model is more sophisticated, and it will consider whether other SVs change lanes when changing lanes. Finally, the level-2 mutual driving model is used as the SV to establish the level-2 training environment, and the competitive and cooperative driving decision models are trained by shaping the reward functions. In particular, the competitive and cooperative decision models are intelligent because they can consider the impact of their actions on SVs when changing lanes.

### C. Training Configurations and Results

TD3 algorithm is used to realize continuous control for car-following driving behavior and human-like social driving policies for the decision-making of the ego vehicle.

The training curve of longitudinal car-following control is illustrated in Fig 6. We also estimated the probability destiny function (PDF) of the car-following gap based on the kernel density estimation (KDE) method, and the distribution is shown in Fig 7. The training curve converges after 90000 training steps. The value of the PDF reaches two peaks at 24m and 27m.

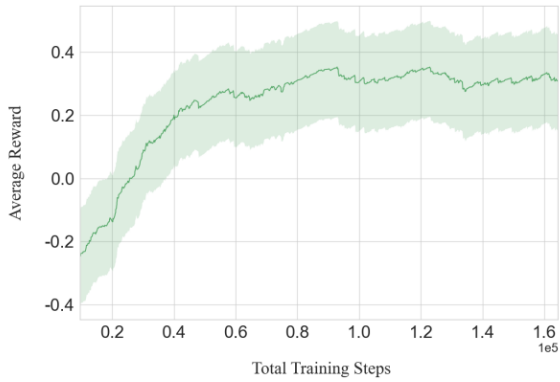

Fig.6 Longitudinal Car-following Control Training Curve

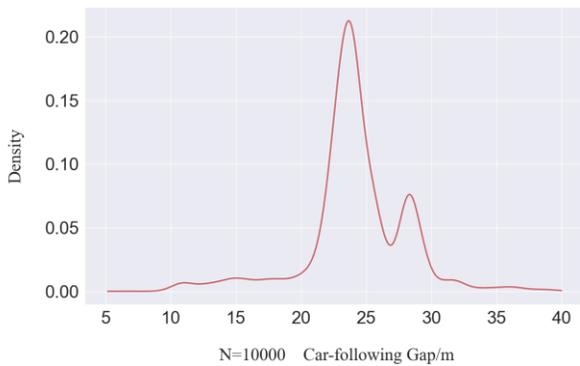

Fig.7 The KDE of Car-following Gap

Four driving policies are generated during the decision training: level-1 mutual driving policy, level-2 mutual driving policy, competitive driving policy, and cooperative driving policy. As shown in Fig.8, the convergence is achieved by around 50000 steps of training.

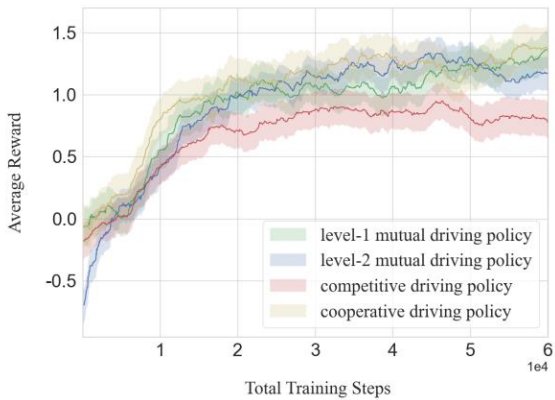

Fig.8 Decision Training Curve

We have recorded a video and selected some episodes of the simulation results to show in Fig. 9. It is noted that all the vehicles that with different driving policies are BVs. We use green to represent cooperative, blue to represent mutual, red to represent competitive, and use the *"idx"* to denote the index. And the ratio of the three types of vehicles is set according to 3:4:3. Moreover, the ratio of the three driving policies can be deployed to the demands of the testing scenario.

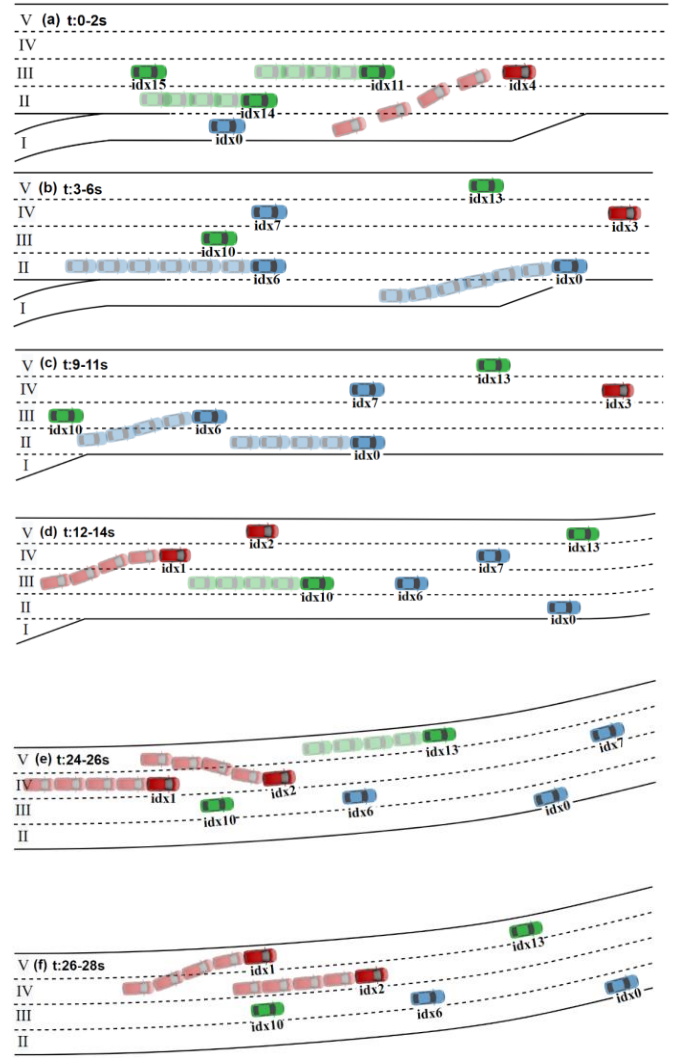

Fig.9 Example of the simulation results consisting of three driving policies. **(a)** The *idx0* with mutual driving policy and the *idx4* with competitive driving policy on-ramp **I** need to merge onto the mainline. The *idx4* first makes two consecutive lane changes to merge onto lane **III**. In addition to gaining higher velocities, the consecutive lane changes reduce the velocities and drivable area of the *idx14* and *idx11*. Both *idx11* and *idx14* decelerate significantly as the *idx4* merges in, showing the setting of cooperation. **(b)** The *idx0* slowly merges onto lane **II** and keeps an appropriate distance from the *idx6* behind,

representing the normal driving behavior of an average human driver. **(c)** As the *idx0* merges onto the mainline at a slower velocity, the *idx6* finds itself still near the *idx0* after slowing down, so the *idx6* changes lanes to the left after observing better driving conditions in lane **III** for driving efficiency and does not continue to reduce velocity. **(d)** When the *idx1* notices that the *idx10* in front of it is going slower, it chooses to make a fast lane change to the left lane **IV**. **(e)** Similarly, due to the slower *idx13*, the *idx2* decides to change lanes to the right lane **IV** while reducing the velocity of the *idx1*. **(f)** Then the *idx1* immediately switches lanes to the left lane **V** to continue driving at a high velocity.

In Fig. 9, the episodes show that the competitive driving policy changes lanes more frequently than the other two driving policies. And the vehicles with this policy could occasionally make aggressive lane changes because they are designed to mimic some aggressive or selfish drivers in the real world. The vehicles with mutual driving policy, on the other hand, usually change lanes slowly to ensure high safety and their driving utilities. Therefore, the lane change behavior of the mutual driving policy is relatively reasonable and normal. Vehicles with cooperative driving policy rarely change lanes as they consider the driving utilities of other vehicles and represent the more conservative drivers in the real world.

## V. EXPERIMENTS AND VALIDATION RESULTS

On the part of the experiment, firstly, the intelligence evaluation framework for the SUTs is introduced. Then, we validated the evolving testing scenario in terms of effectiveness and highlighted the more complex of the evolving testing scenarios compared to other baseline testing scenarios. Lastly, based on the comparisons with statistical NDD (HighD), we demonstrated high fidelity in the evolving testing scenarios.

Three typical driver models, including two rule-based models (Nilsson[48] and MOBIL[49]) and one game theoretic model (Stackelberg[50]), are tested in the evolving scenario, and they are regarded as different SUTs. In terms of testing effectiveness, the testing scenario is valid if the degree of intelligence of the three SUTs can be quantitatively distinguished based on the test results. In terms of complexity validation, an uncertain and safety-critical scenario is considered complex. Besides the evolving scenario, two additional rule-based testing scenarios consisting of Nilsson and MOBIL models, respectively, have also been created. Thus, Stackelberg runs as SUT in three different scenarios, and the higher the scenario complexity score, the more challenging the scenario is for Stackelberg. The relationship between the above two validations is illustrated in Fig. 10.

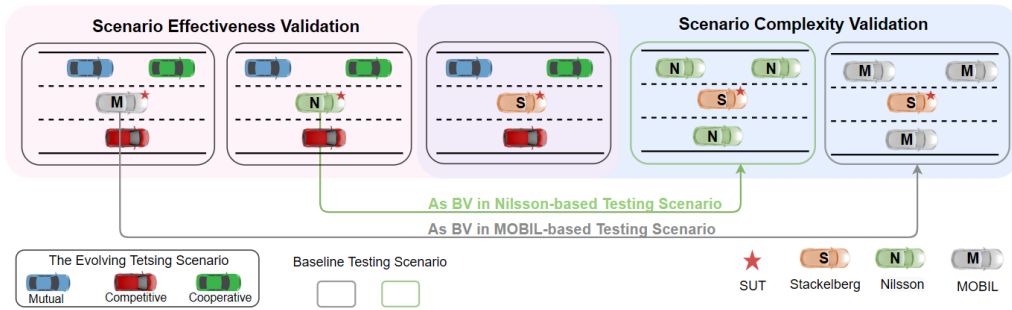

Fig.10 The Validations Design of evolving testing Scenario in Effectiveness and Complexity

### A. Effectiveness Validation of Evolving Testing Scenario

#### 1) Intelligence Evaluation Framework for SUTs

Safety is the key factor in determining whether AVs are industrialized. And driving efficiency is a major factor in determining whether AVs can be better than human drivers. Besides, we propose a new intelligence evaluation metric: interaction utility, which refers to the need for an AV to minimize the impact of its current driving behavior on the driving gains of other vehicles in the scenario when interacting with them. This metric reflects the more advanced intelligence of the BV in terms of dynamic interaction with the other vehicle. Thus, a novel and comprehensive framework for intelligence evaluation is proposed, as shown in Fig.11. Also, the relevant parameters for calculating the three metrics are defined in Table II.

Table II

Parameter Definitions for Intelligence Evaluation

| Parameters | Definition |
|---|---|
| $C$ | Collision times |
| $E$ | Number of exposures to potentially dangerous driving conditions |
| $H$ | Maximum number of vehicles allowed around the SUT |
| $\bar{v}$ | Average driving velocity of the SUT |
| $\bar{h}$ | Average traffic density |
| $\overline{T_l}$ | Average lane-change time |
| $\overline{|v_\sim|}$ | Average velocity fluctuation of BVs in the horizon |
| $\overline{|a_\sim|}$ | Average acceleration fluctuation of BVs in the horizon |
| $\bar{L}$ | During the lane-change process of the SUT, the average lane-change times of BVs in the horizon |

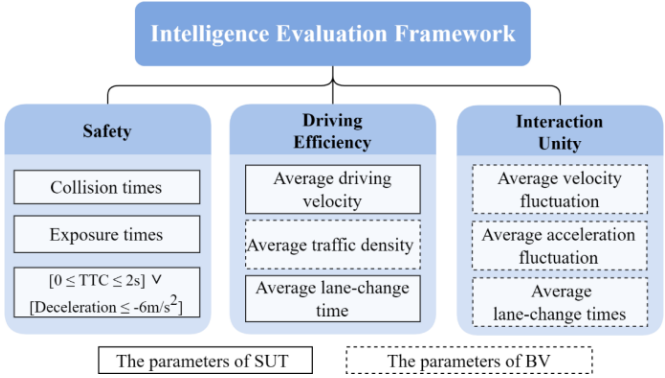

Fig.11 The Intelligence Evaluation Framework for the SUTs

The safety evaluation metric is defined as consisting of the collision times $C$ and the number of exposures $E$ to potentially dangerous driving conditions. Usually, the potentially dangerous driving conditions could be caused by a small time-to-collision (TTC) or a large degree of deceleration. In this paper, when the TTC is less than 2s or the deceleration of SUT is less than -6m/s², the driving condition is convinced to be potentially dangerous. Thus, the safety can be calculated as:

$$\mathbb{I}_s = 0.7 \times \left(1 - \frac{\min(C, C_{max})}{C_{max}}\right) + 0.3 \times e^{-E} \quad (12)$$

We evaluate driving efficiency in terms of three parameters: the average driving velocity, the average lane change time, and the average traffic density in the field of view $\overline{T_l}$. The lane change process is defined as the time elapsed from when the vehicle decides to change lanes to when the center of mass of the vehicle is in another lane. Therefore, the metric of driving efficiency is defined as:

$$\mathbb{I}_e = 0.5 \times \left(\frac{\sqrt{\bar{h}}\bar{v}}{Hv_{max}} + e^{-\frac{\overline{T_l}}{3}}\right) \quad (13)$$

For interaction utility evaluation, we use the five parameters $\overline{|v_\sim|}$, $\overline{|a_\sim|}$, $\Delta v_{max}$, $\Delta a_{max}$ and $\bar{L}$ to represent the impact of the behavior of the SUT on other vehicles in the field of view when a lane change occurs:

$$\mathbb{I}_i = \frac{1}{3} \times \left(2 - \frac{clip(\overline{|v_\sim|}, 0, \Delta v_{max})}{\Delta v_{max}} - \frac{clip(\overline{|a_\sim|}, 0, \Delta a_{max})}{\Delta a_{max}} + e^{-\sqrt{\bar{L}}}\right) \quad (14)$$

Thus, an intelligence evaluation equation is yielded by summarizing the scores of the three metrics, where the $i_s, i_e, i_i$ are the weights.

$$\mathbb{I} = i_s\mathbb{I}_s + i_e\mathbb{I}_e + i_i\mathbb{I}_i \quad (15)$$

2) SUTs

In this paper, only the decision-making system is taken as the tested object. When the tested system gives the policy signal, the car-following behavior and the control logic are used based on the standardized Sumo[51] driver model and PID control. The lane-change models of the three SUTs are briefly described next.

Nilsson model generates a lane-change request by evaluating the utility $U_l$ in each lane $L_l, l \in \mathbb{L}$ where the $\mathbb{L}$ is the set of lanes. The lane with the maximum utility will be chosen as the target lane in the next time step. There are three utility functions in the Nilsson model: the unity $U_{lv}$ of the average travel time, the unity $U_{ltg}$ of the average time gap density, the unity $U_{ld}$ of the remaining travel time. Then the lane utility will be determined by:

$$U_l = w_1 \frac{U_{lv}}{N_{lv}} + w_2 \frac{U_{ltg}}{N_{ltg}} + w_3 \frac{U_{ld}}{N_{ld}} \quad (16)$$

where $w_1, w_2, w_3$ are optional weighting parameters, and $N_{lv}, N_{ltg}, N_{ld}$ are normalizing factors. A detailed explanation of these parameters can be found in[48].

MOBIL model is similar to the Nilsson model in generating a lane-change request. However, the Nilsson model considers all the available lanes in the driving scenario, MOBIL model only considers the nearby lane utilities. The target lane utility is defined as:

$$\widetilde{a_c} - a_c + p[\widetilde{a_n} - a_n + \widetilde{a_o} - a_o] > \Delta a_{th} \quad (17)$$

where the $a_c$ is the acceleration of the ego vehicle in the current lane, $a_o$ is the acceleration of the old follower, the $a_n$ is the acceleration of the new follower. Parameters with the hat ~ indicate the anticipation of the value after the ego vehicle changes to the target lane, the $p$ is the politeness factor which will generate moderate driving policies. The $\Delta a_{th}$ is the threshold to prevent the advantage of lane-change action is only marginal to a lane-keeping action. Further details can be found in[49].

Stackelberg is a typical temporal game-based model in game theory. Three participants are defined in this model, including one leader and two followers. The mechanism remains that the leader will first act, and the other two followers will subsequently take actions based on the action of the leader to maximize the gains. Then, in response to the actions of followers, the leader will take another action to maximize its benefits. The game-based loop will be continued until a Nash Equilibrium is achieved. To simplify the gaming process, only three steps are considered in a game-based loop. The benefit of a specific decision action is evaluated by position utility $U_{pos}$ and negative utility $U_{neg}$, which are defined as:

$$U_{pos} = \begin{cases} \min(d_\Delta, d_v) \text{ if there is a vehicle ahead} \\ d_v \quad\quad\quad\quad\quad otherwise \end{cases} \quad (18)$$

$$U_{neg} = d_\nabla - v_r T - d_{min} \quad (19)$$

where the $d_\Delta$ is the relative distance between the ego vehicle and the nearest BV in the front view, $d_v$ is the maximum visibility distance. $d_\nabla$ and $v_r$ are the relative distance and velocity between the ego vehicle and the nearest BV in the back view, $T$ is the prediction time window, $d_{min}$ is the safety distance of conducting a lane-change action. The three-step Stackelberg gaming indicates that the leader should act to maximize its benefits under the worst circumstance. The optimal decision $\chi_h^*$ should satisfy:

$$\chi_h^* \in \underset{\chi_h}{\text{argmax}} \min_{\chi_1, \chi_2} [U'_{pos} + U'_{neg}] \quad (20)$$

3) Effectiveness of Evolving Testing Scenario

An effective testing scenario is capable of quantifying the intelligence level of different SUTs. We use the intelligence score to evaluate the performance of different SUTs under the same evolving testing scenario. The testing scenario is configured by 20 BVs, where 40% of BVs are generated with competitive driving policy, 30% are mutual driving policy, and the rest are cooperative driving policy. After 3000 simulation steps, the intelligence scores of Nilsson, MOBIL, and Stackelberg are calculated in Table III.

Table III
Intelligence Evaluation of Different SUTs

| SUTs | Safety | Driving Efficiency | Interaction Utility | Intelligence |
|---|---|---|---|---|
| **Stackelberg** | **0.336** | 0.040 | 0.091 | **0.467** |
| **Nilsson** | 0.084 | 0.032 | 0.093 | 0.209 |
| **MOBIL** | 0 | **0.056** | 0.092 | 0.148 |

It can be seen that the Stackelberg achieves the highest score than the other rule-based SUTs. Especially in safety, the Stackelberg only experiences two collisions while the Nilsson and the MOBIL experience several times collisions. For driving efficiency, the number of lane-changes was 125 for Stackelberg, only 20 for MOBIL and 15 for Nilsson, since the lane-change times of Stackelberg are more than the other two SUTs, it is still less driving efficiency than the MOBIL, although its average velocity is the highest. The MOBIL has the highest driving efficiency because its average velocity is higher than the Nilsson and the number of lane-changes is much lower than the Stackelberg. For interaction utility, all SUTs achieve almost the same level.

Fig. 12 shows the distribution of TTC statistics of the SUTs in the simulation testing. Compared with the MOBIL and Nilsson, the Stackelberg encounters the lowest number of dangerous conditions, owing to its TTCs being mainly concentrated around the 20s. The main reason is that Stackelberg makes more game-like decisions and can judge or make lane-changes in advance to eliminate the current dangerous conditions when there are vehicles ahead. In contrast, the MOBIL and Nilsson often need to drive close enough to the vehicle in front before taking the lane-change behaviors.

In order to represent the driving efficiency of vehicles under different decision-making algorithms, we conducted statistics on the average driving velocity of vehicles, and the results are shown in Fig. 13. The average driving velocity of vehicles based on the Stackelberg is almost all concentrated around 30m/s. The average driving velocity of vehicles based on the MOBIL also has a relatively good performance, while the average driving velocity of vehicles based on the Nilsson algorithm is the lowest.

In conclusion, the evolving testing scenario can quantitatively distinguish the intelligence of the game theoretic SUT from the rule-based SUTs.

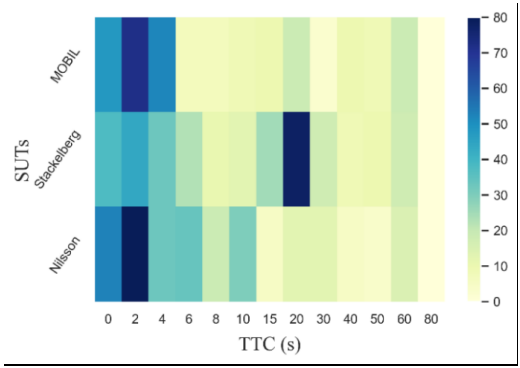

Fig. 12 The TTC Distributions of Different SUTs in Simulation Testing

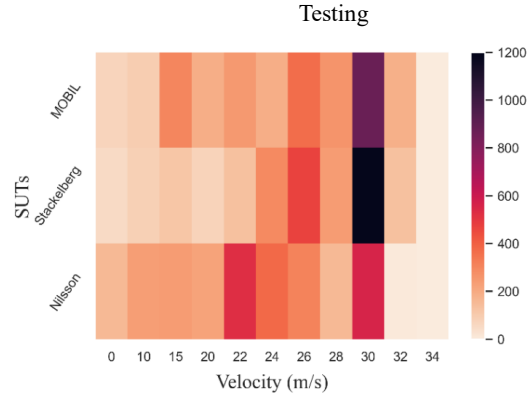

Fig. 13 The Velocity Distributions of Different SUTs in Simulation Testing

*B. Complexity Validation of Evolving Testing Scenario*

A testing scenario is complex means that the interaction of BVs within this scenario is uncertain, as well as the scenario is very safety-critical for the SUT. Therefore, we propose the following three metrics to verify the complexity of the scenario: the times of lane changes ($LC$), the times of acceleration and deceleration maneuvers performed ($AD$), and the number of exposures to potential dangers of the SUT within the scenario ($E$). Therefore, the formula for determining the overall complexity is presented in as follows:

$$\mathbb{C} = 0.4 \times (1 - e^{\frac{-E}{k_1}}) + 0.3(2 - e^{\frac{-LC}{k_2}} - e^{\frac{-AD}{k_3}})$$

where $k_1, k_2\ and\ k_3$ are used to adjust the order of magnitude of the parameter, the values are 25, 8 and 1400 respectively.

Two rule-based testing scenarios are created as baseline scenarios, where the Nilsson and MOBIL are configured as BVs, respectively. After 20 rounds of simulation (about 10,000 steps per round), the average results of 20 BVs and one SUT per 100 steps are listed in Table IV.

Table IV
Safety Evaluation of Stackelberg in Different Testing Scenarios

| BVs and SUT in Scenarios | Lane Change Times ($LC$) | Acceleration and Deceleration Times ($AD$) | Number of Exposures Potential Dangers (SUT) ($E$) | Complexity Score |
|---|---|---|---|---|
| DRL-based | 13 | 1543 | 38 | 0.834 |
| Nilsson | 6 | 1336 | 12 | 0.564 |
| MOBIL | 5 | 1322 | 30 | 0.665 |

For safety simulation testing, the Stackelberg model encounters two collisions in the evolving testing scenario. In contrast, no collision happened in other rule-based testing scenarios. Regarding the number of exposures to potentially dangerous driving conditions, the Stackelberg experienced 38 dangers in the evolving scenario, much higher than the 12 in the Nilsson-based scenario and slightly higher than the 30 in the MOBIL-based scenario. Obviously, the lowest safety score is obtained in the evolving testing scenario, which means that the evolving scenario is safety-critical compared to the other two rule-based scenarios.

*C. Fidelity Validation of Driving Behaviors in Evolving Testing Scenario*

The HighD dataset is chosen for the fidelity validation of the characteristics of car-following and lane-change driving behaviors of BVs in the evolving testing scenario. In order to enhance credibility, more than 0.3 million frames of driving data are compared.

1) Car-following Behavior

For the fidelity validation of car-following driving behaviors, we compared the HighD data with the simulation data of BVs in evolving scenarios, and introduced two metrics which are time headway (THW) and distance headway (DHW) from both temporal and spatial perspectives. We selected one of the HighD data files containing 1,776 vehicles and 15,277 frames with similar vehicle densities to the evolving scenarios for analysis. And we fit the joint PDFs of DHW and velocity, and the joint PDFs of THW and velocity from the two data sources, respectively. The results of the fitting and comparison are shown in Fig.14 and Fig.15. It can be seen that the BVs in the evolving scenario produce the probabilistic distributions that are similar to the naturalistic ones. One of the reasons for the higher distribution of BVs at [26,28] m/s and [80-100] m DHW could be that the competitive driving policy plays an important role. Since this policy prefers to drive at high velocities, it requires a longer distance from the behind vehicle.

In order to more accurately measure the similarity between data distributions from HighD and from evolving scenarios, Jensen–Shannon (JS) divergence is introduced, which is used to solve the problem of asymmetry of Kullback-Leibler (KL) divergence in characterizing the distribution similarity, as the following equations. Since JS divergence is measured using information entropy for two different joint distributions of $p$ and $q$, $JS(p||q)$ reflects the information uncertainty of the distribution $q$ with respect to the distribution $p$. When the two distributions are identical, which means the $q$ does not have any distribution uncertainty based on the distribution $p$, then t $JS(p||q) = 0$. Otherwise, the larger the JS divergence, the lower the similarity (Eq.23). In terms of the characteristics of car-following driving behaviors, the JS divergences between two distributions from different data sources (HighD and the evolving scenario) are listed in Table V.

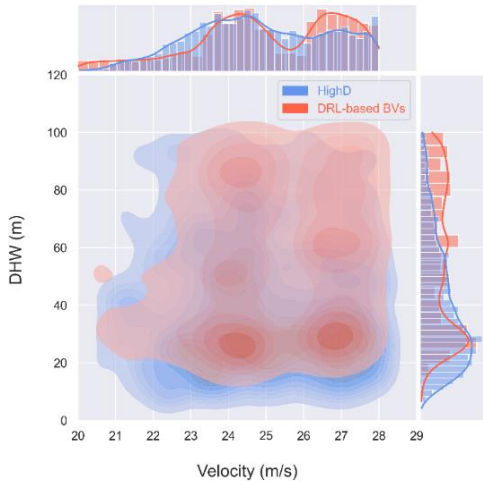

Fig.14 The KDE of Joint PDFs of DHW and Driving Velocity in Car-following

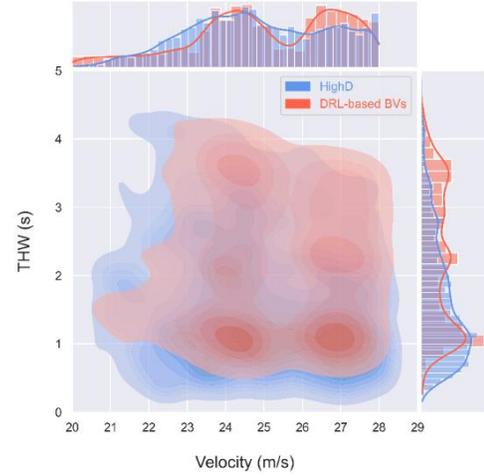

Fig.15 The KDE of Joint PDFs of THW and Driving Velocity in Car-following

$$KL(P \parallel Q) = \sum p(x) log \frac{p(x)}{q(x)} \quad (21)$$

$$KL(Q \parallel P) = \sum q(x) log \frac{p(x)}{q(x)} \quad (22)$$

$$JS(Q \parallel P) = \frac{1}{2} KL\left(P \parallel \frac{1}{2}(P + Q)\right) + \frac{1}{2} KL\left(Q \parallel \frac{1}{2}(P + Q)\right) \quad (23)$$

$$S = \left(1 - JS(HighD \parallel DRL)\right) \times 100\% \quad (24)$$

Table V
JS Divergences Between Data Distributions from Evolving Simulation and HighD Under Car-following Behavior

| Metrics | $JS(HighD||Simulation)$ |
|---|---|
| DHW & Velocity | **0.086** |
| THW & Velocity | 0.007 |

The similarity is more than 90%, which means the characteristics of car-following driving behaviors in the evolving testing scenario are consistent with that in the HighD dataset.

2) Lane-change Behavior

For the fidelity of the characteristics of lane-change driving behaviors, we use the TTC to represent the lane-change motivation at the lane-change moment. Therefore, the fidelity of lane-change action can be validated by comparing the similarity of lane-change motivations between HighD and DRL datasets. We categorize lane-change action into two types: mandatory lane-change and voluntary lane-change. A lane-change action is convinced to be forced if there exists a preceding vehicle (PV) in front of the SUT and the velocity of the PV is smaller than SUT. However, if the SUT conducts a lane-change action when there is no PV in front of it or the velocity of PV is greater than SUT, the lane-change action is deemed voluntary. The KDE of driving velocity under voluntary lane change and the KDE of TTC under mandatory lane change are illustrated relatively in Fig. 16 and 17.

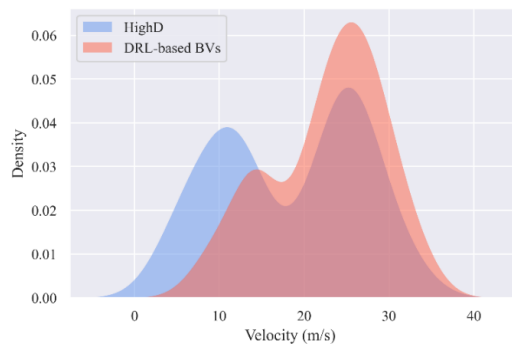

Fig.16 The KDE of Driving Velocity of Voluntary Lane-change Behavior

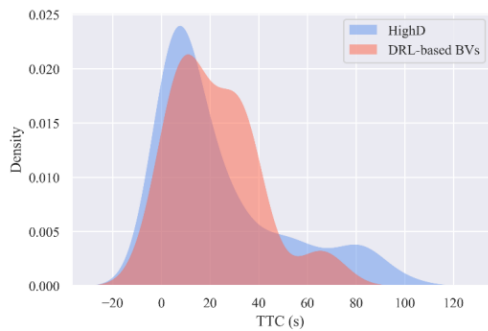

Fig.17 The KDE of TTC of Mandatory Lane-change Behavior

Likewise, the JS divergences between evolving simulation and HighD-based data distributions of different lane-change behaviors are computed in Table VI.

Table VI

JS Divergences Between Data Distributions from Evolving Simulation and HighD Under Lane-change Behavior

| Lane-change Types | $JS(HighD||Simulation)$ |
|---|---|
| Mandatory lane-change behavior | 0.122 |
| Voluntary lane-change behavior | 0.139 |

It can be seen that the similarities of different lane-change behaviors are more than 85%, which proves the fidelity of lane-change motivation. It can be deduced that the evolving testing scenario exhibits a higher degree of consistency with the naturalistic driving environment in terms of car-following and lane-change driving behaviors. As a result, the evolving testing scenario can be considered a high-fidelity representation of real-world driving conditions.

## VI. CONCLUSION

In this paper, by creating human-like social BVs, an evolving testing scenario for the intelligence evaluation of AVs was generated. Firstly, a driver model scheme is designed that incorporates higher-level and lower-level allocation to facilitate realistic interaction. Secondly, different driving policies, including competitive, mutual, and cooperative policies, are shaped by using various reward functions in the TD3 algorithm. Thirdly, a "level-k" game theory approach is employed to train the three driving policies with human-like social interaction. Finally, an intelligence evaluation framework is proposed to rank several SUTs based on their performance under the same evolving testing scenario. The evolving testing scenario is shown to be effective, complex, and high-fidelity in evaluating the intelligence of the SUTs. Future work will focus on implementing the generation method on different road topologies, such as roundabouts and intersections.

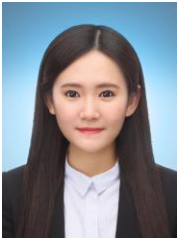
**Yining Ma** received the M.A.Sc degree in Mechanical Engineering from Pusan National University, Pusan, Korea, in 2016 and B.Eng. degree in Vehicle Engineering from the University of Shanghai for Science and Technology, Shanghai, China, in 2014. She is currently working toward the Ph.D. degree in Vehicle Engineering at School of Automotive Studies, Tongji University, Shanghai, China. Her research interests include testing scenario generation and evaluation of autonomous vehicles.

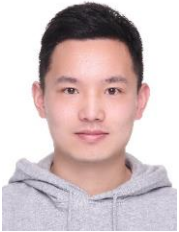
**Wei Jiang** received the B.S. degree in vehicle engineering from Tongji University, Shanghai, China, in 2021. He is currently working toward the M.S. degree in vehicle engineering from the School of Automotive Studies, Tongji University, Shanghai, China. His research interests include intelligent driver model and unknown unsafe scenario generation.

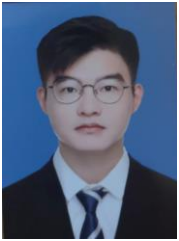
**Lingtong Zhang** received the B.S. degree from Hefei University of Technology, Anhui, China, in 2018, and received the M.S. degree in vehicle engineering from the School of Automotive Studies, Tongji University, Shanghai, China, in 2022. His research interests include reinforcement learning and driver model generation.

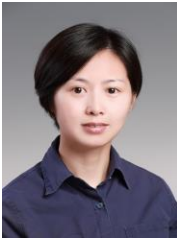
**Junyi Chen** received the B.S., and Ph.D. degrees from the School of Automotive Studies, Tongji University, Shanghai, China, in 2004, and 2009, respectively. She also received the joint Ph.D. degree from the Institut für Mathematische Stochastik, TU Braunschweig, Germany, in 2008. In 2010, she joined the School of Automotive Studies, Tongji University, as a lecturer, till now. Her research topics include safety of autonomous vehicles, driving assistance systems, test and evaluation of autonomous vehicles.

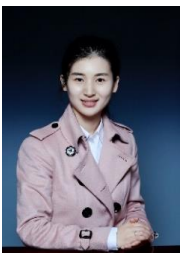
**Hong Wang** is Research Associate Professor at Tsinghua University. She received the Ph.D. degree in Beijing Institute of Technology, Beijing, China, in 2015. From the year 2015 to 2019, she was working as a Research Associate of Mechanical and Mechatronics Engineering with the University of Waterloo. Her research focuses on the safety of the on-board AI algorithm, the safe decision-making for intelligent vehicles, and the test and evaluation of SOTIF. She becomes the IEEE member since the year 2017. She has published over 60 papers on top international journals. Her domestic and foreign academic part-time includes the associate editor for IEEE Transactions on Intelligent Transportation Systems, IEEE Transactions on Vehicular Technology and IEEE Transactions on Intelligent Vehicles, Young Communication Expert of Engineering, lead Guest Editor of Special Issues on Intelligent Safety of IEEE Intelligent Transportation Systems Magazine.

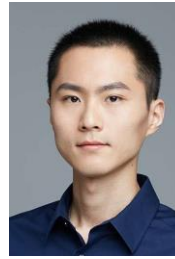
**Chen LV** received the Ph.D. degree from the Department of Automotive Engineering, Tsinghua University, China, in 2016. From 2014 to 2015, he was a joint Ph.D. Researcher with the Department of Electrical Engineering and Computer Science, University of California at Berkeley, Berkeley. He is currently an Assistant Professor with Nanyang Technology University, Singapore. His research interests include advanced vehicle control and intelligence, where he has contributed over 100 papers and obtained 12 granted patents. He received many awards and honors, selectively including the Highly Commended Paper Award of IMechE U.K. in 2012, the NSK Outstanding Mechanical Engineering Paper Award in 2014, the CSAE Outstanding Paper Award in 2015, the Tsinghua University Outstanding Doctoral Thesis Award in 2016, the IV2018 Best Workshop/Special Issue Paper Award, and the Winner of INTERPRET Challenge of NeurIPS 2020 Competition. He serves as an Academic Editor/an Editorial Board Member for IEEE TRANSACTIONS ON INTELLIGENT TRANSPORTATION SYSTEMS and SAE International Journal of Electrified Vehicles and the Guest Editor for IEEE/ASME TRANSACTIONS ON MECHATRONICS and IEEE Intelligent Transportation Systems Magazine.

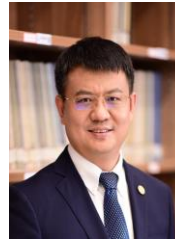
**Xuesong Wang** is a professor of Transportation Engineering at Tongji University, Executive Director of the Joint International Research Laboratory of Transportation Safety, Associate Director of Key Laboratory of Road and Traffic Engineering-China Ministry of Education, and Associate Director of the Engineering Research Center of Road Traffic Safety and Environment Engineering-China Ministry of Education. Dr. Wang's main expertise is in the areas of traffic safety planning, safety evaluation of roadway design, traffic safety management, driving behavior analysis, vehicle active safety. He is a handling editor of Transportation Research Record. He is the Associate Editors of Accident Analysis and Prevention, China Journal of Highway and Transport. Dr. Wang serves as the chair for the safety committee at Shanghai Department of Transportation, the chair for the safety committee at Shanghai Institute of Transportation Engineering. He is an associate editor of Accident Analysis and Prevention. Dr. Wang has published more than 400 papers on academic journals and conferences. Earlier in his career, Dr. Wang has his Ph.D. in Transportation Engineering from the University of Central Florida. In 2015, he was awarded as the outstanding young researcher by the Chinese National Science Foundation.

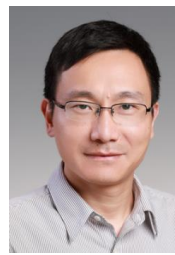
**Lu Xiong** received the B.E., M.E., and the Ph.D. degrees in Vehicle Engineering from the School of Automotive Studies, Tongji University, Shanghai, China, in 1999, 2002, and 2005, respectively. From November 2008 to 2009, he was a Postdoctoral Fellow with the Institute of Automobile Engineering and Vehicle Engines, University of Stuttgart, Germany, with Dr. Jochen Wiedemann. He is currently a Professor with Tongji University. He is also an Executive Director of the Institute of Intelligent and an Associate Director of the Clean Energy Automotive Engineering Center, Tongji University. His research interests include perception, decision and planning, dynamics control and state estimation and testing and evaluation of autonomous vehicles.